\documentclass{aa}  

\usepackage{graphicx}
\usepackage{subfigure}
\usepackage{color}
\usepackage{txfonts}
\begin{document}

   \title{Time-dependent escape of cosmic rays from supernova remnants potentially at the origin of the very-high-energy cosmic-ray gradient of the Galactic center}

   \author{L. Jouvin
          \inst{1}
          \and
          A. Lemi\`ere\inst{2}
          \and
          R. Terrier \inst{2}
          }

   \institute{IFAE, Institut de Física d'Altes Energies, Campus UAB, E-08193 Bellaterra, Spain\\
              \email{ljouvin@ifae.es}
         \and
             APC, AstroParticule et Cosmologie, Universit\'e Paris Diderot, CNRS/IN2P3, CEA/Irfu, \\ Observatoire de Paris, Sorbonne Paris Cit\'e, 10, rue Alice Domon et L\'eonie Duquet, 75205 Paris Cedex 13, France.
             }


 
  \abstract{The distribution of the very-high-energy diffuse emission in the inner 200 pc measured by HE.S.S. indicates the existence of a pronounced cosmic-ray (CR) gradient peaking on the Galactic center (GC). Previous studies have shown that these data are consistent with a scenario in which the CRs are diffused away from a stationary source at the GC. We previously showed that, taking the 3D gas distribution and a realistic distribution of supernova explosions into account, CRs accelerated in supernova remnants (SNR) should account for a large fraction of the GC CRs observed by H.E.S.S.; but the model did not fully reproduce the apparent over-density in the inner 30 pc. 
  
  In this work, we study the time-energy dependent cosmic rays escape from the remnant that is expected to occur when the shock wave decelerates in the surrounding medium. We show that the resulting CR distribution follows the quasi-stationary profile observed by H.E.S.S. more closely. The main signature is that the energy-dependent escape creates a strong dependency of the morphology of the $\gamma$-ray emission with the energy. The existence of this energy dependency should be observable by the Cherenkov Telescope Array.}

\titlerunning{Time-dependent escape of cosmic rays from supernova remnants in the Galactic center}
   \maketitle
%

\section{Introduction}
\label{intro}
H.E.S.S. observations \citep{2006Natur.439..695A, 2016Natur.531..476H, 2018A&A...612A...9H} have revealed the presence of the very-high-energy (VHE) $\gamma$-ray emission (100 GeV - 100 TeV), which is localized along the Galactic plane (GP) in the central 100 pc of our Galaxy. The clear correlation with the matter distribution spread in the central molecular zone (CMZ) indicates that it is mostly due to the interaction of relativistic hadronic cosmic rays (CRs) with the ambient medium. 
A detailed study performed in \citet{2016Natur.531..476H} revealed two observational facts. First, the spectrum extracted in the inner 0.4$^\circ$ shows significant emission up to $\sim$ 40--50 TeV with no indication of a cutoff. This is the signature of the presence of petaelectronvolt particles in the region. Second, the multi-TeV CR density profile deduced from integrated $\gamma$-ray fluxes at various positions shows a marked peak close to the Galactic center (GC) and a decrease at larger distances. This profile is consistent with a 1/r distribution expected in the vicinity of a steady CR accelerator.  
To explain these facts, \citet{2016Natur.531..476H} proposed a scenario in which CR are injected by a stationary source located at the GC, possibly by the supermassive black hole SgrA$^\star$ itself.

While a stationary source and Sgr A$^\star$ can reproduce the H.E.S.S. data well, the contribution expected from CRs accelerated by established particle accelerators such as supernova remnants (SNRs) must be low enough for this to be the main contribution to VHE emission. 
Considering a model in which the CR spectrum is harder in the central kpc of the Galaxy, \citet{2017PhRvL.119c1101G} showed that the total Galactic CR "sea" could significantly contribute to the bulk of this central VHE emission. However, this scenario seems to contradict the actual VHE diffuse emission morphology, which displays a deficit of $\gamma$-ray emission compared to the available matter quantity in the outer part of the CMZ, at $l \approx 1.3^\circ$ \citep{2006Natur.439..695A, 2018A&A...612A...9H}. In addition, their model does not include the specific contribution of SNR in the GC. Even if the supernova (SN) rate in the region suffers from large uncertainties, in \citet{2017MNRAS.467.4622J} we show that SNR should release large power in the form of energetic particles. Assuming a CR acceleration efficiency of only 2\% of the kinetic energy released at the SN explosion, the CRs released by the SNe explosion in the GC can already reproduce the total $\gamma$-ray flux observed with H.E.S.S..

Yet, the maximal energy reached by the CRs from the acceleration of isolated SNRs is still under debate. Some models have proposed that CR can reach petaelectronvolt energy in core-collapse SNe with dense circumstellar winds \citep{2010ApJ...718...31P} or type Ibc SNe with transrelativistic shocks \citep{2013ApJ...776...46E}.  We note that a large fraction of massive stars reside in the massive and dense young stellar clusters in the region (namely, the central disk, the Arches cluster, and Quintuplet cluster) so that collective effects, such as colliding shock flows, could play a role. Colliding show flows have been shown to be very efficient in increasing the SNe acceleration efficiency \citep{2018AdSpR..62.2764B, 2019AdSpR..64.2439B}.  The role of collective effects has recently been emphasized by \citet{2019NatAs...3..561A} comparing the CR profiles around several massive star clusters and the GC. Following \citet{2017MNRAS.467.4622J}, in this work we do not enter into these theoretical considerations and we assume that those sources are capable of accelerating particles up to petaelectronvolt energy.

In \citet{2017MNRAS.467.4622J}, we built a 3D model of CR injection and propagation in the CMZ to obtain the $\gamma$-ray emission produced by the CRs interacting with ambient matter. This model uses the 3D matter distribution obtained by \citet{2004MNRAS.349.1167S} as well as a spatial and temporal distribution of SNRs based on the star formation rate (SFR) observed in the region as a whole and inside the main massive star clusters. The abundance of these impulsive accelerators creates a sustained CR injection in the region and they should not be neglected in any model of $\gamma$-ray emission. The realistic SNe spatial distribution considered in this model, with a concentration in the central massive star clusters, also creates a CR gradient toward the GC. Only the very central excess seemed difficult to reproduce by the accelerated CRs from SNRs alone.

  However, this model was too simplistic since it assumed a complete burst-like injection of all particles from the accelerator at the same time. There are theoretical  \citep[e.g.,][]{2003A&A...403....1P} and observational evidences that the most energetic particles escape from the remnants during their first phases of evolution, while lower-energy particles stay trapped over much longer periods. Such runaway particles might explain the GeV emission of a molecular cloud in the surrounding of the SNR W44 \citep{2012ApJ...749L..35U} as well as the teraelectronvolt emission of the molecular clouds in the vicinity of the SN SNR W28 \citep{2008A&A...481..401A}. During the adiabatic phase of the remnant evolution, the shock wave decelerates and the CRs are progressively released in the ISM as the maximal particle energy the shock can sustain decreases. The low-energy CRs remain in the remnant longer than the high-energy CRs leading to an energy-dependent CR injection in the ISM \citep{2005A&A...429..755P, 2009MNRAS.396.1629G}. 
  
  In this work, we improve our model with a more physical CR escape scenario from the SNR, following the approach of \citet{2009MNRAS.396.1629G}, which depends on the deceleration of the shock wave with the time and the CR energy. Because the low-energy CRs remain in the SNR longer than the recurrence time between two SNe explosions, this should create a quasi-stationary CR distribution at least in the lower-energy part of the spectrum. We can expect that the global morphology of the $\gamma$-ray emission becomes similar to that produced by a stationary source at the GC at low energy, while at the highest energies it should follow the flatter profile expected in the impulsive case. In this scenario, we should expect a pronounced variation of the emission morphology with the energy. This variation could be used to distinguish the different scenarios with the next generation of Cherenkov telescopes of the CTA observatory \citep{2011ExA....32..193A}.

  In section \ref{model}, we first note the main physical assumptions of the 3D model we built in \citet{2017MNRAS.467.4622J} and we describe how the time-energy-dependent escape from the SNR is modeled in this work. Then in section \ref{results}, we investigate the impact on the VHE $\gamma$-ray morphology of combining a real distribution of those sources in the region with a more realistic CR escape from the SNR. We compare with H.E.S.S. observations and the burst-like scenario and we explore the impact of the physical parameters of our model on the VHE $\gamma$-ray emission. We finally discuss in section \ref{cta_predict} the perspectives with CTA comparing the H.E.S.S. and CTA observations of this central region.

\section{Time-energy-dependent 3D model of CR injection and gamma-ray production}
\label{model}
In this section, we describe the time-energy-dependent CR escape from the SNR, which we include in the 3D model of CR injection and propagation in the GC developed in \citet{2017MNRAS.467.4622J}. We first describe the main physical aspects of this model. 

\subsection{Context of the 3D model}
\label{article_precedent}
The main elements necessary to build a 3D model of the gamma-ray emission are the spatial distribution of sources, in this work assumed to be SNRs, and the 3D distribution of matter.

The presence of the three most massive stellar clusters of the Galaxy in the central 30 pc of the GC---the Quintuplet, the Arches, and the central disk surrounding SgrA$^\star$---indicates the necessity to take into account the concentration of the SNe at their position. Since one-third of the massive stars detected in the GC are located outside of these three massive starburst clusters, suggesting the existence of isolated high-mass star formation \citep{2010ApJ...725..188M}, a uniform distribution of the SNe throughout the region also has to be considered. We generated the SNe with a global recurrence time of 2500 years based on the central value estimated by \citet{2011MNRAS.413..763C} in the region. The final spatial distribution is composed of two components: first, a uniform component, where the SNe are uniformly distributed in a cylinder of 150 pc radius and 10 pc height; and, second, a concentration of the sources in the central massive star clusters taking into account a SN rate at their position that is compatible with that determined from physical observations. As discussed in \cite{2017MNRAS.467.4622J}, the Arches cluster, which has an estimated age of 2-3 Myrs \citep{1999ApJ...525..750F},
 seems too young to have experienced any SN. Its position is excluded. The other important aspect for the resulting morphology of the emission is the 3D matter distribution that is hard to build in the GC region owing to a lack of information on gas kinematics. For a latitude of 0$^\circ$, we based our distribution on the work of \citet{2004MNRAS.349.1167S}, which allows us to trace the dense molecular clouds of the region as well as a widespread high-temperature and lower-density diffuse molecular component that could represent 30\% of the total mass of the CMZ \citep{2001A&A...365..174R}. To get the complete 3D distribution, we assumed an exponential decay along the Galactic latitude following \citet{2007A&A...467..611F}.

Knowing that the kinetic energy released from a SN explosion $E_{SN}$ is $\rm 10^{51} \, erg$, the $\gamma$-ray luminosity observed by H.E.S.S. appears to be several orders of magnitude lower than the available power of these sources \citep{2011MNRAS.413..763C}. Using a simple one zone steady-state model, we demonstrated that the diffusion is much more competitive than advection for the CRs to escape the region at these very high energies, assuming a diffusion coefficient close to the ISM value. The CR transport equation used to determine the CR propagation is thus modeled by the simple isotropic diffusion equation. As discussed in \citet{2017MNRAS.467.4622J}, we considered a power-law diffusion coefficient, $D=D_0 {\left(E/10 \ \rm{TeV}\right)}^{\rm d}$, which has a typical value $d=0.3$ and a diffusion coefficient value at 10 TeV of $2\times 10^{29}$ $\rm cm^2s^{-1}$ close to the ISM value. The 3D Green function, obtained assuming a CR density equal to zero at $r=+\infty$, gives the solution of the diffusion equation for an impulsive accelerator and can be used to model the CR injection by the SNe. The CR injection spectrum follows a power law, $N_o E^{-a}\delta(r-r_0)\delta(t-t_0)$, with a spectral index $a$ equal to 2 since this value is typical for diffusive shock acceleration. Whereas in the previous study we considered only impulsive injection, in this work we modeled the CR time-energy-dependent escape from the SNR.

\subsection{Time-energy-dependent CR escape from SNRs}
\label{phase_sedov}

Particles acceleration occurs in the first two evolution phases of the SNR: the free expansion phase in which the ejecta evolves almost freely (approximately a few hundred years) followed by the adiabatic phase in which the shock wave deceleration starts. The highest energy reached should occur at the end of the free expansion phase \citep{2003A&A...403....1P}. As a result of the progressive deceleration of the shock wave, during the adiabatic
phase the CRs are injected into the ISM at different times, depending on their energy, owing to different confinement times in the remnant. Therefore the maximum energy reached  by the CRs in this phase decreases but its evolution with time is not yet well known. We followed the approach of \citet{2009MNRAS.396.1629G}, in which the time dependency of the energy loss of the protons follow a power law of index  $-1/k$ with the energy. Within this model, the solution of the simple diffusion equation that we considered in this study (Sect. \ref{article_precedent}) is analytical. \footnote{We note however that more complete models, such as \cite{2011ApJ...731...87E}, via the Monte Carlo technique, predict more complex behaviors.} The CR final differential spectrum at a distance $r$ from the source and at a time $t$ after the source explosion is given by

\begin{center}
\begin{multline}
\dfrac{dN}{dE_p} (r,t,E_p) =\dfrac{N_0 {E_p}^{-a}}{{(4\pi D(t-\chi (E)))}^{\frac{3}{2}}} \\  \times \exp \left( \frac{-r^2}{4D(t-\chi(E))}\right) \, \rm{TeV^\mathrm{-1}\mathrm m^\mathrm{-3},}
\label{stefano}
\end{multline}
\end{center} 

where $\chi (E)= t_{Sedov} \, \left(\dfrac{E}{E_{max}}\right)^{-1/k}$, where $E_{max}$ is the maximal energy at which the CRs are accelerated during the Sedov phase. The quantity $\chi (E)$ is the time after the explosion at which a proton of energy $E$ escapes from the SNR. Following \citet{2009MNRAS.396.1629G}, $E_{max}$ = 5 PeV at the early epoch of the Sedov phase ($t_{Sedov}$=200 years) and $E_{min}$ = 1 GeV at the late epoch of the Sedov phase ($5 \times 10^4$ years). With those parameters fixed, we considered the same CR energy-loss dependency as \citet{2009MNRAS.396.1629G} obtained to reproduce well the theoretical modelings of CR acceleration and escape from the SNRs \citep{2003A&A...403....1P}. Assuming this energy-loss scenario, the time $\chi (E)$ at which the CRs escape from the SNR after the explosion is represented in Fig. 1. The low-energy protons remain confined longer in the SNR and on a larger timescale that the SN recurrence time, whereas those at higher energy have already escaped between each SN explosion. As we describe in the next section, this differential effect combined with the real spatial distribution of the SNe produces a $\gamma$-ray emission profile similar to a single stationary accelerator at the center.
 
 \begin{figure}
\centering
\subfigure{\includegraphics[width=\columnwidth]{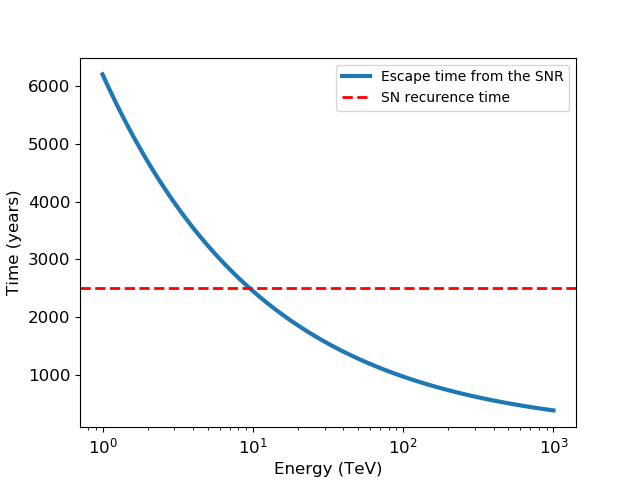}}
\caption{Escape time of CRs from SNR after explosion as a function of their energy (blue line). The SN recurrence time in the region is indicated as a dashed red line.}
\label{escape_time}
\end{figure}

 The previous estimation is performed for typical conditions in the ISM with a density around $\sim$ 1 $\rm{cm^3}$. The time at which the SNR enters in the radiative phase depends on the medium conditions where the SNR is located. The transition to the radiative phase is faster in a dense medium since the shock slows down more quickly. The maximal energy of accelerated CRs in the adiabatic phase decreases with time. Depending on the duration of the adiabatic phase, the energy-loss dependency of the protons changes. We discuss these consequences in the section \ref{param_influence}.

Following our previous study (section \ref{article_precedent}), we simulated the injection and propagation of CRs from 1 TeV to 1 PeV in a 3D box of size $500 \, \rm {pc} \times 500\, \rm {pc} \times 50 \, \rm {pc}$ centered on the GC. We assumed, as a hypothesis of our model, that the maximal energy reached by the CRs in the acceleration process is 1 PeV. We then focussed on the observational consequences of this hypothesis on the observed VHE emission. We do not consider sources $<1 \, \rm{kyr}$ since any younger SNR has been observed within the GC. We only keep sources $<$ 100 kyr since for older events the CR density becomes negligible. The table \ref{model_param} contains the different physical parameters used in our model. These are identical to those used in \citet{2017MNRAS.467.4622J}. In this work, we study the CR escape time after the SN explosion, which is energy dependent; this is the only additional parameter of the model with respect to \citet{2017MNRAS.467.4622J}. It is represented in function of the energy on Fig. \ref{escape_time} and the values at 1 TeV and 10 TeV are given in Table 1. We used the same analytical shape described in \citet{2017MNRAS.467.4622J} to compute the total $\gamma$-ray flux produced by the pp interaction. The $\gamma$-ray emissivity obtained in each pixel of the cube is then integrated along the line of sight to compare with the H.E.S.S. data. 

\begin{table}
\caption{Physical parameter values used in this 3D model of CR injection and propagation in the GC. These values are the same as in \citet{2017MNRAS.467.4622J}, except for the CR escape time that had been added to the initial model.}
\label{model_param}
\begin{tabular}{|p{4cm}|p{4cm}|}
        \hline
   Model parameters & Values\\
   \hline
   Proton spectral index & 2 \\
   $E_{max}$ of the injected proton spectrum & 1 PeV \\
   Box size & $500 \, \mathrm pc \times 500\, \mathrm pc \times 50 \, \mathrm pc$ \\
   Total gas mass  & $4\times 10^7$ $\rm{M_{\odot}}$  \\
   $D_0$ (10 TeV) & $2\times 10^{29}$ $\rm cm^2s^{-1}$ \\
   Spectral index of the diffusion coefficient (d) & 0.3 \\
   $E_{SN}$ & $10^{51}$ $\rm erg$ \\
   SN recurrence time & 2500 yrs \\  
   Escape time after the SN explosion at 1 TeV & 6000 years\\  
   Escape time after the SN explosion at 10 TeV & 2500 years\\  
\hline 
\end{tabular}
\end{table}

\section{Results and discussion}
\label{results}

We compared the model predictions of a time-energy-dependent CR injection from the SNRs to the H.E.S.S. data and to the simple burst-like injection scenario studied in \citet{2017MNRAS.467.4622J}. For both scenarios, we generated 100 spatial and temporal random SNe distribution. We then used the median and the dispersion around the median of the realizations of these distributions to compare with the H.E.S.S. observations. The dispersion represents 50\% of the realizations for each longitude bin independently.

We focus mainly our comparisons on the morphology of the distributions using the $\gamma$-ray and CR longitude profiles extracted in the region (Fig. \ref{CR_profile} and Fig. \ref{profil_gamma_normalised}). In \citet{2017MNRAS.467.4622J}, to avoid overproducing the $\gamma$-ray spectrum observed by H.E.S.S., we concluded that the CR acceleration efficiency in the SNR has to be very low, roughly 2\% of the kinetic energy released at the SN explosion. The predicted spectrum obtained by adding the time-dependent CR escape from the SNR is compatible with the burst-like scenario and does not add any constraint. Unlike a stationary scenario, we can expect some spectral variability throughout the region resulting from  the fast escape from the GC at higher energy. However, the variation of the spectral index between both scenarios that we estimate with our model ($\sim$ 0.15) is hardly detectable by the current air atmospheric telescopes owing to their statistical and systematical uncertainties. In the near future, a study of the morphology of the $\gamma$-ray emission in several energy bands will be a powerful tool to distinguish the scenarios.

\subsection{Peaked CR density profile}

\begin{figure}
\centering
\subfigure{\includegraphics[width=\columnwidth]{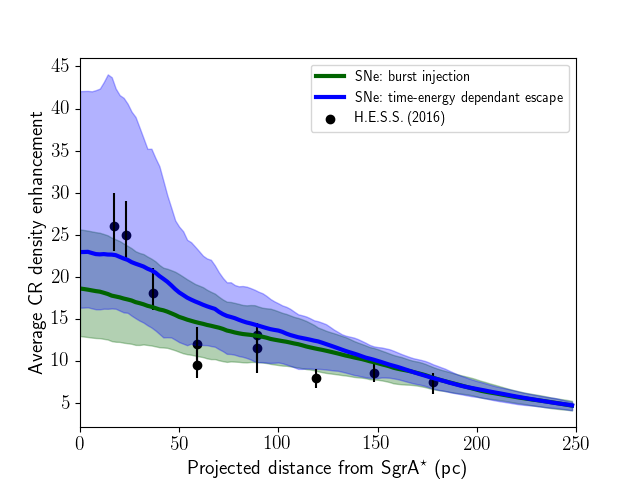}}
\caption{Average CR enhancement with the projected distance to the GC extracted from H.E.S.S. data \citep{2016Natur.531..476H} in black and predicted by our 3D model of the SNe in the region considering a burst-like injection (green, \citep{2017MNRAS.467.4622J}) and a time-energy-dependent escape from the SNR (blue). These profiles are the mean of the profiles for Galactic latitudes (i.e., $\rvert b \rvert <0.1^\circ$). The solid lines represent the median of the 100 spatial and temporal SN distributions and the colored regions the dispersion around this median.}
\label{CR_profile}
\end{figure}

Figure \ref{CR_profile} represents the CR density profile as a function of the projected distance from the central source SgrA$^\star$ for both scenarios, compared to that extracted from data by the H.E.S.S. collaboration \citep{2016Natur.531..476H}. They deduced the CR density within three ring regions and seven circular regions of radius 0.1$^\circ$ distributed along the GP by converting the VHE $\gamma$-ray luminosity extracted in these regions with the total molecular mass obtained from the CS lines for each of them. We used the same approach as in \citet{2017MNRAS.467.4622J} \footnote{A small mistake was written in the text and formula for the determination of the average energy density of CR from our model that we correct here. We determined the average energy density of CR above 10 TeV along the line of sight by weighting the CR energy density predicted in each pixel of the 3D box by the matter quantity in each pixel: $n_{CR}(x,z) = \frac{\int_{y} w_{CR}(x,y,z) \times n(x,y,z) \, dy }{\int_{y} n(x,y,z) \, dy}$, where $n_{CR}$ is the CR energy density in Galactic latitude ($z$) and Galactic longitude ($x$), $n(x,y,z)$ the matter density in each pixel of the 3D box, $w_{CR}(x,y,z)$ the CR energy density, and $y$ the direction along the line of sight.} to extract the CR density from our 3D model. For the burst-like scenario \citep{2017MNRAS.467.4622J}, we previously showed that it seemed unlikely for the impulsive accelerators alone to reproduce the very central CR excess, even if it is compatible with the error bars for certain SNe realizations. By adding the more realistic time-energy-dependent CR escape from the SNR, the CR density profile appears more peaked toward the center since the main effect is to lead to a quasi-stationary injection at lower energy (Fig. \ref{escape_time}). The median of the different realizations is now compatible with the H.E.S.S. data points including in the central part. The dispersion around the median increases compared to the burst-like scenario. Since the low-energy CRs do not diffuse far from their emission point, the youngest explosion in the central region has more impact on the resulting profile. As explained in \citet{2017MNRAS.467.4622J}, in our approach the total mass of molecular gas is factored out when computing predicted CR profiles. Conversely, data points extracted in \citet{2016Natur.531..476H} depend on the actual H$_2$ mass and must  account for uncertainties on the CS to H$_2$ conversion factor. Taking into account those uncertainties, both the burst-like or time-dependent injections are consistent with the H.E.S.S. CR profiles. However the time-dependent injection clearly makes the emission more peaked toward the center owing to a pseudo-stationary injection at the center for the low-energy part of the spectrum.

Taking into account nonstationary effect, a realistic 3D spatial matter and SN distributions (concentrated in the central part of the GC but not as close to SgrA$^\star$), we showed that SNe create a smooth CR gradient toward the center \citep{2017MNRAS.467.4622J}. Adding this time-energy dependent CR escape from the SNR creates a strong peak in the CR density at the GC. The spectrum predicted in the region is perfectly compatible with the H.E.S.S. $\gamma$-ray spectrum that led to the indication for petaelectronvolt particles in the region. Since the CR profile extracted from the $\gamma$-ray emission is extracted over a large and integrated energy range, a stationary injection at the GC at petaelectronvolt energies is not necessary. The pseudo-stationary injection created at the low-energy part of the spectrum in our scenario also creates a peaked CR distribution toward the GC that is compatible with the H.E.S.S. results. With this approach, we show that the CR is not required to be accelerated constantly very near SgrA$^\star$ to reproduce the observed peaked CR profile, as proposed by \citet{2016Natur.531..476H}.

As detailed in the next section, the evolution of the morphology of the $\gamma$-ray emission will be the key observable to be able to distinguish between a single stationary source at the center and multiple accelerators.
 
\subsection{Energy-dependent morphology of the $\gamma$-ray emission}
\label{morpho_dep}

\begin{figure}
\centering
\subfigure{\includegraphics[width=\columnwidth]{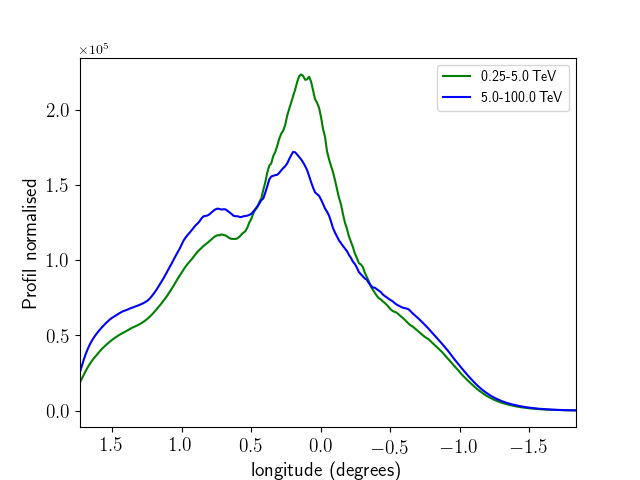}}
\caption{Median of the expected normalized VHE $\gamma$-ray profile along the Galactic longitude obtained from 100 SN temporal and spatial distributions taking into account the time-energy-dependent escape from the SNR, after integrating along the line of sight and the Galactic latitude b, $\rvert b \rvert <0.1^\circ$. The profiles in the two energy bands, 0.25-5 TeV and 5-100 TeV,  are normalized.}
\label{profil_gamma_normalised}
\end{figure}

In the time-energy-dependent escape scenario, a permanent CR injection in the central region of the GC occurs into the low-energy part of the spectrum. The $\gamma$-ray profile in the low-energy band (Figure \ref{profil_gamma_normalised}) seems therefore more similar to that produced by a stationary source at the GC \citep{2017MNRAS.467.4622J}. On the other hand, at higher energies, the CRs have already escaped between two SNe explosions. Therefore, the profile appears flatter and more closely follows the dense gas distribution since the CR population is more uniform and similar to that of a burst-like injection (Fig. \ref{profil_gamma_normalised}).

This significant variation of the emission morphology with the energy is critical to distinguish this model from the scenario of a stationary injection at the center from SgrA$^\star$ that produces a morphology stable across the energy range.  At higher energies, the central peak should disappear from the $\gamma$-ray profile. It means that if the peak persists across the energy range, a scenario closer to a stationary source should be favored. As we discuss in section \ref{cta_predict}, the variation in energy of the spatial features of this emission will be precisely characterized with the future generation of Cherenkov telescopes that will be available with the CTA observatory.

\subsection{Influence of the parameters}
\label{param_influence}
In \citet{2017MNRAS.467.4622J} we discussed the fact that the CR acceleration efficiency in the SNR is highly dependent on the choice of some physical model parameters such as the SN rate, the total molecular mass, and the diffusion coefficient value. Exploring the variance of these parameters, we concluded that this efficiency should lie between 0.5\% to 20\% to reproduce the data. In addition to the influence on the available energy power, parameters such as  the diffusion coefficient or the duration of the adiabatic phase have an impact on the morphology of the emission. We briefly discuss their influence below.

The diffusion coefficient value we adopt in this work lies in the lower range of different estimations by \citet{2014PhRvD..89h3007G, 2011ApJ...729..106T} based on local CRs measurements.  By increasing the value, the CRs escape faster from the region and, for the low-energy part of the spectrum especially, they are less confined in the central part of the GC. Therefore the predicted CR gradient toward the GC could decrease. However even by considering a diffusion coefficient higher by a factor of 10, the CR profile remains compatible with the H.E.S.S. data points. The variation of the  morphology of the $\gamma$-ray emission in the different energy bands becomes even stronger in the case of a higher diffusion coefficient value, since the difference in the CR escape time from the GC region between the low and high energies increases. The CR acceleration efficiency needed in this latest scenario becomes closer to the more standard value considered for SNRs (i.e., roughly 10\%). The magnetic field observed in the GC is approximately poloidal on average in the diffuse inter-cloud medium \citep{2014arXiv1406.7859M}. We also considered an anisotropic diffusion coefficient, using higher value in the direction perpendicular to the GP and found that this does not significantly change the conclusions.

The CR escape from the SNR occurs during the adiabatic phase of the remnant expansion. Since the temperature and the density in the GC are higher than those in the general ISM, the shock in the adiabatic phase could become supersonic more rapidly. Therefore the transition to the radiative phase occurs faster and the time spent by the CRs in the remnants decreases. The CR energy losses during the adiabatic phase is described by a power law of index 1/k (Sect. \ref{phase_sedov}) that controls the stationarity of the emission. The limit $k \rightarrow \infty$ is the scenario of an impulsive time injection independent of the energy. By increasing the value of $k$, since the CRs time spent in the SNR decreases, the stationarity of the injection becomes even more limited on the low-energy part of the spectrum. The differential effect in energy also becomes less important. Therefore the variation of the morphology of the emission across the energy range decreases. If the CR escape time from the SNR is not too low compared to the SNe recurrence time, the $\gamma$-ray profile remains close to a stationary one at low energy. Moreover, if a SN event recently occurred at the center, the very central CR excess is still reproduced by this scenario of multiple accelerators.

\section{Perspective with CTA}
\label{cta_predict}

 The GC is one of the key science projects of the observatory \citep{2017arXiv170907997C} with more than 700 hours that will be dedicated to observe this region. Whereas observing the predicted spectrum variability of the $\gamma$-ray emission is likely to be challenging for this next generation of telescopes, it will be possible to create much more detailed morphological maps and to determine precisely if the central $\gamma$-ray excess persists toward the highest energies since we will have access to many more morphological details on the VHE emission of the CMZ. In particular thanks to the Small Size Telescope (SST), the effective area at high energy will increase.

In this section, we compare H.E.S.S. and CTA observations of this central region by simulating GC observations using the software \textit{Gammapy} (version v0.17 \footnote{\url{https://docs.gammapy.org/0.17/}}) \citep{2017ICRC...35..766D}, an open-source Python package for $\gamma$-ray astronomy, which provides tools to simulate and analyze the $\gamma$-ray sky for imaging atmospheric Cherenkov telescopes.

To simulate the observed count map in different energy bins we could obtain from a real observation, we used a given 3D model in space and energy of the $\gamma$-ray emission and a given set of instrument response functions \footnote{\url{https://gamma-astro-data-formats.readthedocs.io/en/latest/irfs/irf\_components/index.html}} (IRF) 
composed of the effective area, energy resolution, point spread function, and background model. For the model, we used the 3D $\gamma$-ray emission cube predicted by our model directly (this work, Sect. \ref{model}). To simulate GC CTA observations, we will use the public CTA IRF \footnote{\url{http://www.cta-observatory.org/science/cta-performance/}}. For H.E.S.S., we used the IRFs delivered in the first H.E.S.S. public data release in the CTA public fits format \footnote{\url{https://gamma-astro-data-formats.readthedocs.io/en/latest/}} \citep{2018arXiv181004516H}. Since the CTA IRF are only available for a zenith of 20 degree, to make the comparison we selected the same observational condition for the H.E.S.S. data. We simulated a dataset of 250 hours of observations (similar to the total observation time published by H.E.S.S. in \citet{2018A&A...612A...9H}. For each observation, the pointing position is located at 0.7 $^\circ$ from GC source. For both telescopes, we generated 100 synthetic datasets with \textit{Gammapy} by generating random count maps from a Poisson distribution taking into account the expected residual hadronic background and the physical model.

\begin{figure}
\centering
\subfigure[H.E.S.S. (250 hours)]{\includegraphics[width=\columnwidth]{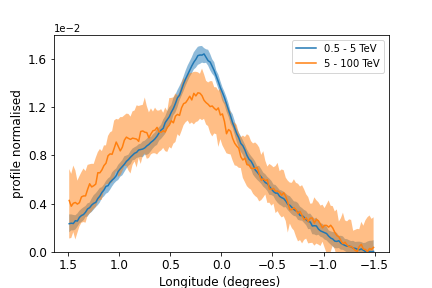}}
\subfigure[CTA (250 hours)]{\includegraphics[width=\columnwidth]{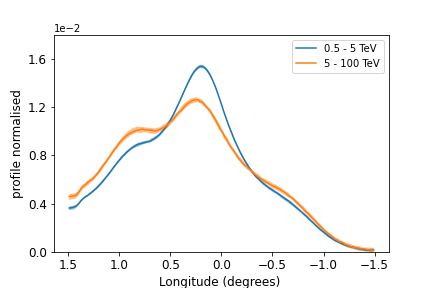}}
\caption[Caption for LOF]{Expected $\gamma$-ray profile along the Galactic longitude from the simulation of 100 GC datasets of 250 hours between 0.5 - 5 TeV (blue) and 5 - 100 TeV (orange). On the top those profiles are simulated using H.E.S.S. IRFs and on the bottom using CTA IRFs (version prod3b-v2, \url{http://www.cta-observatory.org/science/cta-performance/}). The solid lines represent the median of the 100 simulated datasets and the colored region the dispersion around the median. The dispersion contains 50\% of the realizations in each longitude bin independently.}
\label{profil_gamma_HESS_CTA}
\end{figure}

The expected $\gamma$-ray profile along the Galactic longitude from those simulations for H.E.S.S and CTA is represented in Fig. 4 in two energy bands: 0.5-5 TeV and 5-100 TeV. 
We can see the variation of the profile morphology in the two energy bands for H.E.S.S. and CTA simulations, but it appears that the statistical errors with H.E.S.S. at high energy are too high to enable us to distinguish any variation between the low- and high-energy profiles (see Fig. \ref{profil_gamma_HESS_CTA}.a). On the other hand, this simulation clearly demonstrates that CTA will be able to measure such a morphology variation with energy thanks to the SST and the strong increase of the statistic in the high-energy range  (Fig. \ref{profil_gamma_HESS_CTA}.b). In the near future,  CTA observations will therefore establish a clear constraint on the origin of this VHE diffuse emission and on the CR acceleration mechanism in the CMZ. 

\section{Conclusions}

SNRs are natural candidates for Galactic CR acceleration. With a 3D model of CR injection and propagation in the GC, we already showed that the SNe alone can reproduce the total VHE $\gamma$-ray flux of the diffuse emission, by considering typical values for the SN rate, the diffusion coefficient, the total mass of the matter in the region, and a low CR acceleration efficiency of approximately 2\% \citep{2017MNRAS.467.4622J}. Besides, we showed that SNRs, concentrated at the position of the massive stellar clusters, the Quintuplet, and the central disk, creates a pronounced CR gradient but can only marginally reproduce the data obtained by the H.E.S.S. collaboration in the inner 30 pc \citep{2016Natur.531..476H, 2018A&A...612A...9H}. Yet, we considered only a burst-like injection in which all the CRs are emitted at the same time in the ISM independently of their energy.

The deceleration of the shock wave in the adiabatic phase of the SNR evolution leads to a progressive energy-dependent CR escape in the surrounding medium. To add this more realistic time-energy-dependent escape from the SNR, we followed the work of \citet{2009MNRAS.396.1629G} who assume that the time dependency of the proton energy loss in this phase follows a power law. With this approach, the low-energy CRs remain confined a longer time in the remnant than the high energy CRs. At lower energies, the CRs do not escape from the central region between two successive SNe explosions. Therefore a pseudo-stationary injection occurs where SNe explosions are frequent, as around the GC. As a result, the predicted CR density gradient is stronger toward the GC. No additional VHE component at the center is required to be in agreement with the CR density profile extracted from the H.E.S.S. observation.

Furthermore, the differential effect created by this time-energy-dependent CR escape from the remnant implies a strong variation of the morphology of the $\gamma$-ray emission with the energy range. On the other hand, in the case of a single stationary source at the center, the morphology is expected to be stable. With the better sensitivity and angular resolution of the new CTA Observatory, the statistic available will allow for a detailed morphology study as a function of the energy. A clear answer on its variation, and therefore on its origin, could be given. We also note that CTA might be able to resolve some of the active accelerators currently injecting CRs into the CMZ.

\section*{Acknowledgements}
We acknowledge the Fran\c cois Arago Centre at the APC laboratory in Paris and the Port of Scientific Information (PIC) at the campus UAB next to the laboratory IFAE in Barcelona to provide the resources necessary to compute the simulations presented in this work.
This research has made use of the CTA instrument response functions provided by the CTA Consortium and Observatory, see http://www.cta-observatory.org/science/cta-performance/ (version prod3b-v2) for more details.

\bibliographystyle{aa}
\bibliography{biblio.bib} 

\end{document}